\documentclass[pra,aps,showpacs,twocolumn]{revtex4}
\usepackage{epsfig}
\usepackage{color}


\begin{document} 
\title{Thermalization in a one-dimensional integrable system} 

\author{Pjotrs Grisins$^1$ and Igor E. Mazets$^{1,2}$}
\affiliation{$^1$Vienna Center for Quantum Science and Technology, Atominstitut, TU Wien, 1020 Vienna, Austria \\
$^2$Ioffe Physico-Technical Institute, 194021 St.Peterburg, Russia } 

\begin{abstract} 
We present numerical results demonstrating the possibility of thermalization of 
single-particle observables in a one-dimensional  system,
which is integrable in both the quantum and 
classical (mean-field) descriptions (a quasicondensate of ultracold, weakly interacting 
bosonic atoms are studied as a definite example). We find that certain 
initial conditions admit the  
relaxation of  single-particle observables 
to the equilibrium state reasonably close to that corresponding to 
the Bose-Einstein thermal distribution of Bogoliubov 
quasiparticles. \\
\end{abstract} 
\maketitle 

\section{Introduction}
A one-dimensional (1D) system of identical bosons with contact interactions is known to be 
integrable since Lieb and Liniger have solved analytically the corresponding 
quantum problem by means of Bethe ansatz \cite{LL}. In the weakly interacting limit, this system 
can be described in the mean-field approximation by the Gross-Pitaevskii  
equation (GPE), also known as the nonlinear Schr\"odinger equation (NLSE). Zakharov and Shabat \cite{ZS73}
have demonstrated that the NLSE with defocusing nonlinearity 
(which corresponds to the repulsive interactions between particles)   
is integrable by the inverse scattering transform (see  
\cite{ISTM} for a general review of the inverse scattering transform method). 
Since the number of integrals of motion 
in an integrable system equals to the number of degrees of freedom (infinite in the continuous 
mean-field description \cite{ZS73} or equal to the number of particles in the quantum Lieb-Liniger model \cite{LL}), 
one might expect that the finally attained equilibrium state must still bear signatures of the initial conditions. 

One-dimensional bosonic systems have been experimentally implemented with ultra-cold atoms on atom chips \cite{H1,H2}, with the 
radial trapping frequency being $\sim 10^3$ times higher than the longitudinal one. The ultracold 
degenerate atomic system (quasicondensate, i.e. a system describable by a macroscopic wave function with a 
fluctuating local phase) was in the 1D regime since both the temperature and the mean 
interaction energy per atom were well below the energy interval between the ground and the first excited states of 
the radial motion. The fact that the static and dynamic correlation properties of these systems were in a very 
good agreement with the Bose-Einstein  equilibrium distribution of quasiparticles seemed to be in contradiction 
with the system integrability and called for explanation. To explain the observed relaxation of single- and two-particle distribution functions 
for the elementary excitations (Bogoliubov quasiparticles) to the Bose-Einstein equilibrium, a mechanism of 
integrability breakdown via three-body effective collisions involving virtual excitations of the radial degrees of 
freedom has been proposed \cite{M089}. 

In the present paper we numerically show the existence of a certain case of nonequilibrium initial conditions 
of the GPE, which provide a very fast  relaxation of the simplest (single-particle) observables   
to an equilibrium state very close to thermal equilibrium, despite the integrability of the problem.

Some  indications of thermalization in 1D bosonic systems have been obtained in 
numerical simulations of various physical processes in quasicondensates, such as the   
subexponential decay of coherence between coherently split quasicondensates 
\cite{Stim1}, soliton formation in a 1D bosonic system  in the course of (quasi)condensation  \cite{DZW}, 
in-trap density fluctuations \cite{Proukakis1},   wave chaos \cite{wchaos}, and condensate formation after 
the addition of a dimple to a weak harmonic longitudinal confinement of a 1D ultracold atomic gas \cite{Proukakis2}. 
However, a systematic study of thermalization of the GPE solution in the course of time evolution was lacking up to now. 
Even \cite{Burnett1}, where thermalization of the GPE solution with the initial conditions 
corresponding to the high-temperature limit has been numerically obtained, states that formal and  
systematic understanding of the problem is still incomplete. 
We fill this gap, at least to a certain extent, with our present study.  

We also have to draw a clear distinction between our approach and that of Rigol \textit{et al.} \cite{Rigol1}, 
who theoretically studied dephasing in a quantum system of hard-core bosons on a lattice, 
prepared initially in a coherent superposition of   
eigenstates, and its relaxation to a generalized Gibbs 
(fully constrained) equilibrium. 
Our aim is to demonstrate that a weakly interacting 1D degenerate bosonic gas 
can approach, in the course of its evolution, a state that is reasonably close to the 
conventional thermal Bose-Einstein equilibrium.

\section{Numerical approach}
We solve the GPE 
\begin{equation} 
i\hbar \frac \partial {\partial t} \Psi (x,t)=-\frac {\hbar ^2}{2m}
\frac {\partial ^2}{\partial x^2} \Psi (x,t)+g|\Psi (x,t)|^2 \Psi (x,t) ,
\label{eq:1} 
\end{equation} 
where $\Psi (x,t)$ is a classical complex field representing a 
quasicondensate of atoms with mass $m$ and $g$ is 
the effective coupling constant in one dimension (we assume $g>0$). 
The interaction strength is characterized by the Lieb-Liniger parameter \cite{LL}  
$\gamma =mg/(\hbar ^2\bar{n})\equiv (\bar{n}\xi )^{-2}$, where $\xi $ is the quasicondensate healing length 
and $\bar{n}\equiv \langle |\Psi (x,t)|^2 \rangle $ is the mean 1D number density. We consider the weak interaction 
limit $\gamma \ll 1$. We assume periodic boundary conditions for $\Psi (x,t)$, with the period $L$ 
being long enough to ensure the loss of correlations over the half period: $\langle \Psi ^*(x,t)\Psi (x+L/2,t)
\rangle \ll \bar{n}$. The 
angle brackets denote here averaging over the ensemble of realizations. 
For each realization  
the initial conditions are prepared   
in a manner similar to the truncated Wigner approach \cite{TW} but taking into account  
thermal fluctuations only (cf. Ref. \cite{Stim1}). 
We express the macroscopic order parameters in terms of the phase $\phi $ and 
density $\delta n$ fluctuations: $\Psi = (\bar{n}+\delta n)^{1/2} e^{i\phi }$. The initial (at $t=0$) 
fluctuations  are expanded into plane waves as  
\begin{eqnarray} 
\delta n(x,0)&=&2\sqrt{\bar{n}/L} \sum _{k\neq 0} \beta _k\sqrt{\eta _k/\epsilon _k}\cos (kx +\varpi _k) ,\nonumber \\ 
\phi (x,0)&=&(1/\sqrt{ \bar{n}L }) \sum _{k\neq 0}\beta _k\sqrt{\epsilon _k/\eta _k}\sin (kx +\varpi _k) ,
\label{eq:2} 
\end{eqnarray} 
where $\epsilon _k=\sqrt{\eta _k(\eta _k +2g\bar{n})}$ is the energy of the elementary (Bogoliubov) excitation 
with the momentum $\hbar k$ and $\eta _k=(\hbar k)^2/(2m)$. The real numbers $\beta _k$ and $\varpi _k$ have the 
meaning of the scaled amplitude and the offset of the thermally excited elementary wave with the momentum $\hbar k$ 
at $t=0$. The values of $\varpi _k$ are taken as (pseudo)random numbers uniformly distributed between 0 and $2\pi $. 
Each ensemble of realizations is also characterized by a distribution of the $\beta _k$ values with 
$\langle \beta _k^2 \rangle $ being equal to the main number $ {\cal N}_0(k)$ of elementary excitation quanta 
(quasiparticles) in the given mode \cite{prim1}. In  equilibrium at the temperature $T$ the populations of the 
bosonic quasiparticle modes are given by   
${\cal N}_\mathrm{BE}(k,T)=\{ \exp [\epsilon _k/(k_\mathrm{B}T)]-1\} ^{-1}$. 

The use of the classical field (GPE) approach is justified, as it has been shown  \cite{SMSM} 
that the noise and correlations in an 
atomic quasicondensate are dominated by thermal (classical) fluctuations under experimentally feasible 
conditions, and the observation of quantum noise is a challenging task that can be solved in 
particular regimes by means of involved experimental tools \cite{Armijo}.

\begin{figure}[b]

\centerline{\epsfig{file=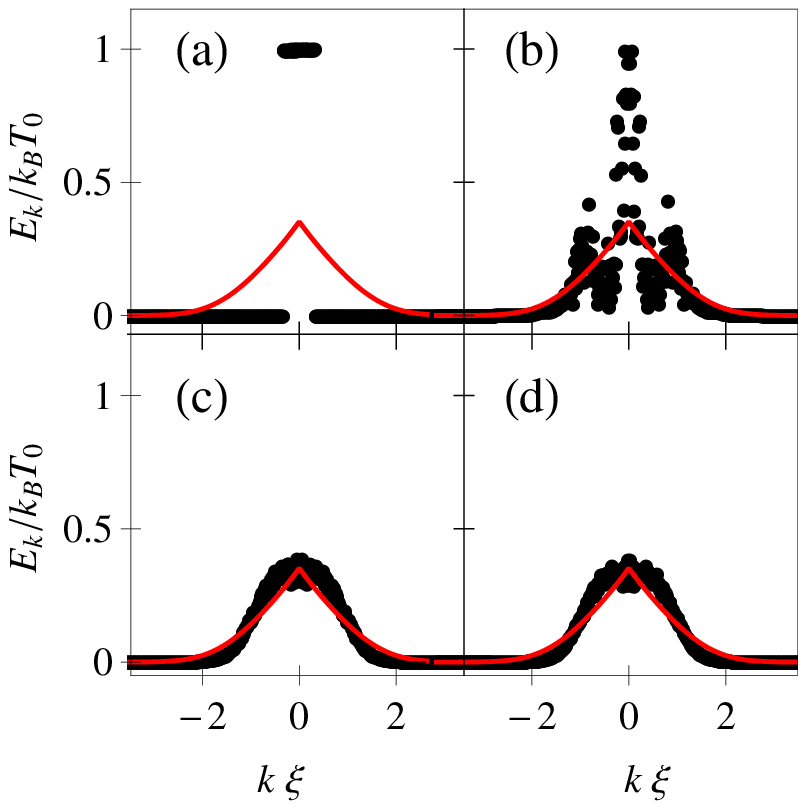,width=0.93\columnwidth}} 

\begin{caption} 
{ \label{f:1}  (Color online) 
Dots: mean energy per mode (scaled to $k_\mathrm{B}T_0 $ with 
$k_\mathrm{B}T_0 = 2\, g\bar{n}$) as the function of wavenumber $k$ (scaled to $\xi $) for 
the dimensionless time $g\bar{n}t/\hbar =$ (a) 0, (b) 50, (c) 2850, and (d) 5750. The Lieb-Liniger 
parameter $\gamma = 5\times 10^{-3}$, $k_0\xi =0.33$. Solid line: mean energy per mode 
$\epsilon _k{\cal N}(k,T_\mathrm{eq} )$ for the equilibrium state,  $T_\mathrm{eq}=0.35\, T_0$; 
see Eq. (\ref{eq:3}). The data are averaged over 200 realizations. 
Units on the axes in this figure and the subsequent figures are dimensionless. }
\end{caption} 

\end{figure} 

\begin{figure}[b]

\centerline{\epsfig{file=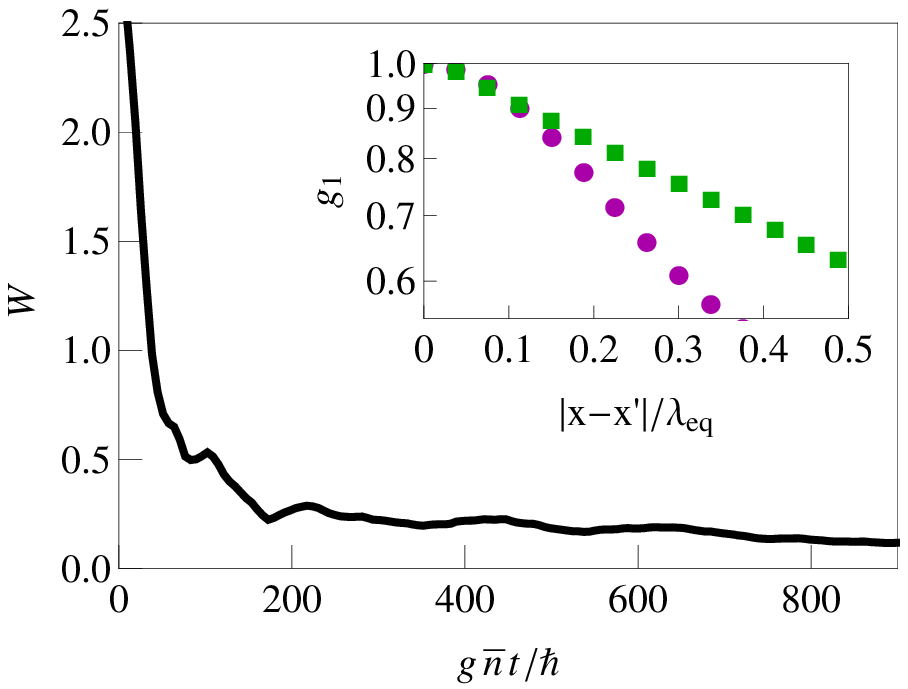,width=0.93\columnwidth}} 

\begin{caption} 
{ \label{f:therm-measure}  (Color online) 
Numerically obtained energy-weighted squared deviation of the quasiparticle 
distribution from the  Bose-Einstein thermal equilibrium as a function of 
time. The initial energy distribution and other parameters are the same as in 
Fig.~\ref{f:1}(a). The inset shows the numerically calculated first-order correlation function $g_1(x-x^\prime )$ 
(shown on the logarithmic scale) 
for the dimensionless time $g\bar{n}t/\hbar =$ 0 (circles) and 6000 (squares). The distance is scaled to 
$\lambda _\mathrm{eq} $.  } 
\end{caption} 

\end{figure}

To integrate Eq. (\ref{eq:1}), we used the fourth-order time-splitting Fourier spectral method \cite{Thalhammer},
which is rather similar to that used in Ref. \cite{Stim1}.

We have found a set of 
examples of solutions of Eq. (\ref{eq:1}) that demonstrate quite a good degree of thermalization. 
Efficient thermalization has been observed in the cases of initial population of     Bogoliubov 
modes within a certain momentum band around $k=0$ [for simplicity, we assume ${\cal N}_0(k)={\cal N}_0(-k)$], 
with the bandwidth being narrow enough to ensure the phononic nature of these excitations, $|k|\xi \lesssim 1$. 
In Fig.~{\ref{f:1}}, we present our results of numerical 
integration of Eq.\, (\ref{eq:1})  for the initial conditions corresponding to the truncated classical 
distribution, parametrized by the effective temperature $T_0$ and the cutoff momentum $\hbar k_0$, i.e., for 
${\cal N}_0(k)$ being equal to $k_\mathrm{B}T_0/\epsilon _k$ for $|k|<k_0$ and zero otherwise. 
For the sake of convenience, in Fig.~{\ref{f:1}} 
we plot  the mean energy per mode $E_k=\epsilon _k{\cal N}(k)$, which does not diverge at 
$k\rightarrow 0$, 
in contrast to the time-dependent population distribution ${\cal N}(k)$. Practically, $E_k$ can be calculated 
by averaging over the ensemble of realizations the energy stored in the given mode:
\begin{equation} 
E_k=\left \langle \frac {m}{2} \bar{n} |v_k|^2 + \left( \frac {\hbar ^2k^2}{8m\bar{n}}+ \frac {g}{2}\right)
|\delta n_k|^2 \right \rangle , 
\label{eka} 
\end{equation}  
where $\delta n_k$ and $v_k$ are the Fourier transforms of the density $\delta n(x,t)$ and velocity 
$v(x,t)=(\hbar /m)\partial \phi /\partial x$ fluctuations. 

Elementary excitations at different momenta are found to be uncorrelated for all propagation times, i.e., 
$\langle \delta n_{k^\prime }\delta n_k^* \rangle = \langle |\delta n_k|^2 \rangle \delta _{k\, k^\prime }$ and  
$\langle v_{k^\prime }v_k^* \rangle = \langle |v_k|^2 \rangle \delta _{k\, k^\prime }$, 
as expected for a thermal equilibrium state. 

The energy distribution  approaches  its equilibrium, which is quite close to the thermal Bose-Einstein 
distribution. The main difference is that the former is flat 
at $k\rightarrow 0$ and the latter has a cusp there. 
The equivalent temperature $T_\mathrm{eq}$ of the corresponding Bose-Einstein 
thermal  distribution is determined from the energy conservation \cite{prim2}: 
\begin{equation} 
\sum _{k\neq 0} \epsilon _k{\cal N}_0(k) = 
\sum _{k\neq 0} \epsilon _k{\cal N}_\mathrm{BE}(k,T_\mathrm{eq}). 
\label{eq:3} 
\end{equation} 
Note that for a weakly interacting  1D system of $^{87}$Rb atoms with the parameters as in Fig.~\ref{f:1} 
the time unit  $\hbar /(g\bar{n})\approx 0.1$ ms. 

To check our numerical method, we performed the following tests. First, we checked the isospectrality 
of the (generalized) Lax operator of the inverse  scattering problem \cite{ZS73,ISTM}. 
We calculated the spectrum of the linear differential operator $\left( \begin{array}{cc} 
i\partial /\partial \bar{x} \, & q \\ q^*  & -i\partial /\partial \bar{x} \end{array} \right) $, where 
$\bar{x}=x/\xi $ and $q= \bar{n}^{-1/2} \Psi (x,t)$, by substituting  the 
numerically obtained solution for $\Psi (x,t)$ at different times  
and comparing the result to the spectrum that corresponds to the initial condition $\Psi (x,0)$. 
The spectrum of the Lax operator 
has been found to be time independent with a high accuracy. The maximum relative shift  of an 
eigenvalue over more than 100 realizations was about $10^{-7}$ for a numerical grid consisting of 1024 points in $x$.
 
Then we checked the time independence of the numerical values of the integrals of motion of Eq. (\ref{eq:1}). 
The first three of them are (up to a numerical factor) the particle number, the total momentum, and the total 
energy of the system. Other integrals of motion can be calculated using the recurrent formula \cite{ZS73}. 
We found that they are conserved with high accuracy, with the relative error being of order of $10^{-11}$ for the first
integral of motion (the number of particles) and of order of $10^{-4}$ for the 15th integral of motion.

Following Ref. \cite{wchaos}, we estimated  the numerical error through the fidelity, defined as 
${\cal F}=\left|1 - (\bar{n}L)^{-1}\int _0^Ldx\, \Psi ^*(x,0) \Psi_\mathrm{fb}(x,t ,-t) \right| $, 
where $\Psi_\mathrm{fb}(x,t,-t)$ is the numerical solution of the GPE with the initial condition $\Psi (x,0)$ 
first propagated forward in time (up to time $t$) and then propagated backward over the same time interval. 
We obtained  ${\cal F}\sim 10^{-8}$ for the propagation times $t$ as long as $ 10^3 \, \hbar /(g\bar{n})$, which is 
sufficient for the establishment of equilibrium, with the spatial grid consisting of 512 points.

\begin{figure}[b]

\centerline{\epsfig{file=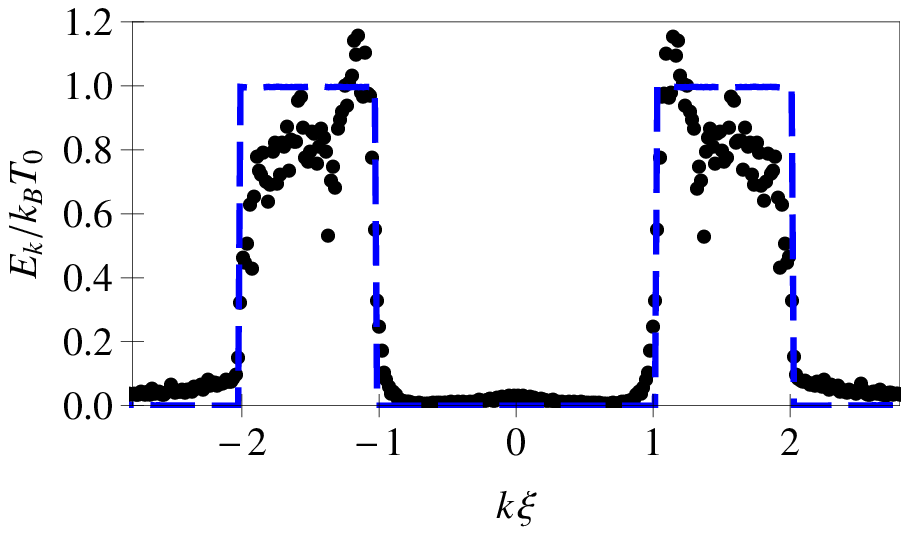,width=0.93\columnwidth}} 

\begin{caption} 
{ \label{f:energies-split}  (Color online) 
Dots: mean energy per mode (scaled to $k_\mathrm{B}T_0 $ with 
$k_\mathrm{B}T_0 = 0.66\, g\bar{n}$) as a function of wave number $k$ (scaled to $\xi $) for 
the dimensionless time $g\bar{n}t/\hbar =2\times 10^4$, 
which is long enough to provide equilibration. The Lieb-Liniger 
parameter $\gamma = 5\times 10^{-3}$, $k_1\xi =1.0$, $k_2\xi =2.0$.  Note the closeness 
of the equilibrium states to the initial energy distribution (dashed line). 
}
\end{caption} 

\end{figure}

We found that our method converges if the grid contains more than 200 points for $L\approx 400\, \xi $. 
A coarse grid (about 100 points) yields a numerical 
artifact: any initial distributions rapidly smears out to the ``classical-like" flat distribution of the energy over 
modes, i.e. to $E_k\approx $\, const for all momenta $-\frac \pi {\Delta x} <k< \frac \pi {\Delta x}$ resolvable by 
the grid with the step $ {\Delta x}$. 

To quantify relaxation of the system toward its equilibrium, we introduce the   measure  
\begin{equation} 
{W}=\frac { \sum _{k\neq 0} \left \{ {\epsilon _k} 
[{\cal N}(k) -{\cal N}_\mathrm{BE}(k,T_\mathrm{eq})]\right \} ^2} {\sum _{k\neq 0}  [ 
{\epsilon _k}{\cal N}_\mathrm{BE}(k,T_\mathrm{eq})]^2} ,
\label{aa} 
\end{equation} 
which has a meaning of the normalized energy-weighted squared deviation of the quasiparticle 
distribution from the  Bose-Einstein thermal equilibrium. 
For the parameters of Figs. \ref{f:1} and ~\ref{f:therm-measure}, with $T_0=150$ nK, the thermalization time is
$\tau _\mathrm{eq}\sim 20$ ms. If we change $T_0$ to 50 nK 
and $k_0\xi $ to 1, then   $\tau _\mathrm{eq}$ decreases by an order of magnitude. 
Note, that the obtained thermalization time $\tau _\mathrm{eq}$ is always shorter 
than the time needed for a sound wave 
to traverse the distance $L$. Therefore the  thermalization observed in our 
simulations is a local physical effect, which is not related to specific boundary conditions. 
The thermalization time $\tau _\mathrm{eq}$ should not be confused with the time $\tau _\mathrm{d}\sim 
m\lambda _\mathrm{T}^2/\hbar $ \cite{Stim1}, where $\lambda _T=2\hbar ^2\bar{n}/(mk_\mathrm{B}T)$, of dephasing between two 1D quasicondensates initially 
prepared in thermal-like states with strongly mutually correlated fluctuations.  

\section{Discussion and conclusions}
Therefore we found numerically an example of the GPE solution that relaxes  
toward a state with practically measurable noise and correlation 
properties \cite{DRBK} well describable by a thermal Bose-Einstein ensemble of 
quasiparticles. As an illustration,  in the inset in 
Fig.~\ref{f:therm-measure} we plot the numerically calculated first-order 
correlation function $g_1(x-x^\prime )=\langle \Psi ^*(x^\prime ,t)\Psi (x ,t)\rangle /\bar{n}$ for 
$t=0$ and for $t$ large 
enough to provide equilibration \cite{prim3}. 
We see that this correlation function finally approaches  the exponential form 
$g_1(x-x^\prime )=\exp (-| x-x^\prime | /\lambda _T)$, predicted for the thermal equilibrium \cite{Popov}, 
with $T\approx T_\mathrm{eq}$ [the distance in the inset to Fig. \ref{f:energies-split}  is scaled to 
$\lambda _\mathrm{eq}= 2\hbar ^2\bar{n}/(mk_\mathrm{B} T_\mathrm{eq})$]. 

\begin{figure}[t] 

\centerline{\epsfig{file=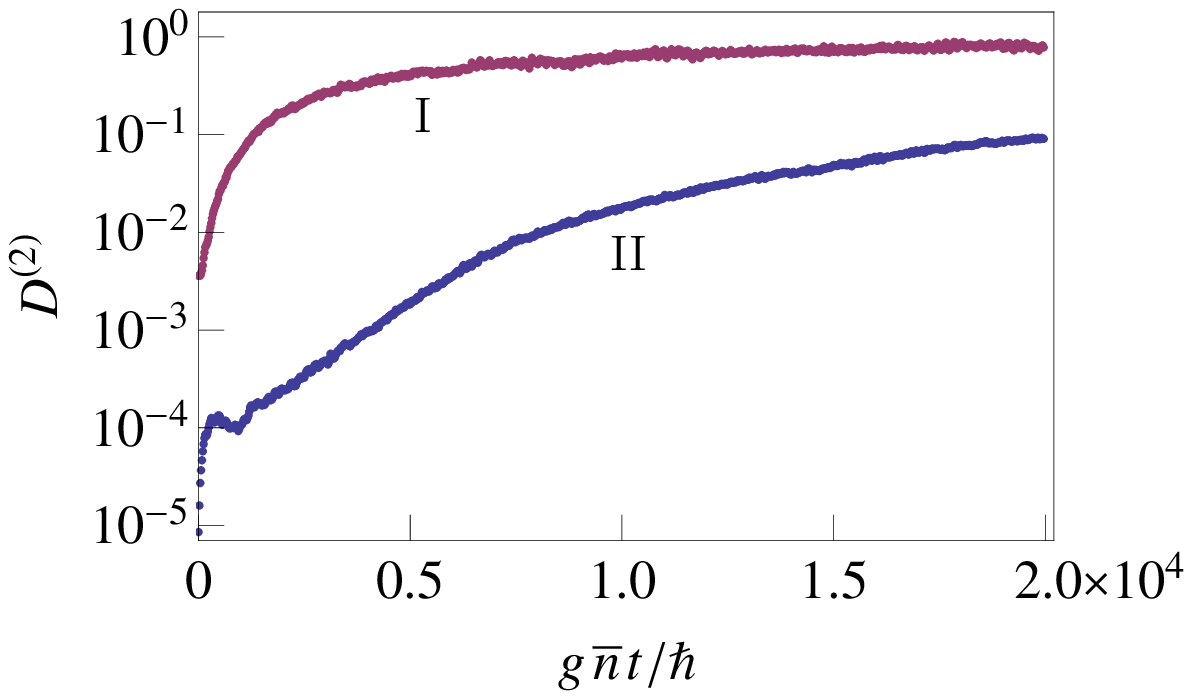,width=0.93\columnwidth}}   

\begin{caption} 
{\label{f:4}  
(color online) Distance $D^{(2)}[\psi _1,\psi _2]$ (on the logarithmic scale, dimensionless) 
as a function of scaled time for the parameters of Fig. \ref{f:1} (I, upper curve) and 
Fig. \ref{f:energies-split} (II, lower curve). }
\end{caption}
\end{figure}

Not every initial distribution relaxes toward the Bose-Einstein thermal equilibrium. For example, 
if  there are initially two oppositely propagating bunches of particle-like elementary excitations well separated 
in the momentum space, an equilibrium state very far from ${\cal N}_\mathrm{BE}(k,T_\mathrm{eq})$ is established, 
as seen from Fig. \ref{f:energies-split}, where we assume $\epsilon _k{\cal N}_0(k)$ 
to be equal to $k_\mathrm{B}T_0$ for $k_1<|k|<k_2$ and zero otherwise ($k_1\gtrsim \xi ^{-1}$). 
This behavior can be viewed as a conspicuous example of  relaxation 
toward the fully constrained equilibrium \cite{Rigol1} in the weakly interacting  case. 

To elucidate the qualitative difference between the cases shown in Figs. \ref{f:1} and 
\ref{f:energies-split}, we calculate the time dependence of the distance 
$ 
D^{(2)}[\psi _1,\psi _2]=(2\bar{n}L)^{-1}\int _0^Ldx\, |\psi _1(x,t)-\psi _2(x,t)|^2 
$ 
between two solutions $\psi _1,\psi _2$ of the GPE, which are very close at $t=0$. 
As we can see from Fig. \ref{f:4}, if phononic modes are initially populated, $D^{(2)}$ grows exponentially and  
saturates at the unity level (corresponding to 
the total loss of correlations at $t\rightarrow \infty $), thus signifying the chaotic regime. 
If only particle-like modes are 
initially populated, then $D^{(2)}$ grows very slowly and stays well below 1 at all experimentally relevant 
times (hence, the chaotic behavior is practically not observed in that case). 

To conclude, we numerically observed thermalization in a 1D quasicondensate, 
i.e. in an ultracold atomic system described 
by the NLSE with a cubic repulsive nonlinearity, if only phononic modes are populated initially. 
The correctness of the numerical solution has been checked via the criteria of the Lax operator isospectrality, 
conservation of the integrals of motion, and fidelity. Such a series of tests prevents the 
possible numerical artifacts that may occur  
in  the split-step method \cite{s20}. 
Although the thermalization is not complete, 
experimentally measurable correlations are expected to be well described by the thermal equilibrium 
of bosonic elementary excitations. Our findings are  
in good agreement with the high efficiency of the evaporative cooling of ultracold atomic gases on the atom chips 
deeply in the 1D regime \cite{H1,H2} (our work on numerical modeling of evaporative 
cooling of ultracold bosonic atoms in elongated traps is in progress). 
On the other hand, to provide full thermalization of 
nonequilibrium ensembles of particle-like excitations, like the one displayed in Fig.~\ref{f:energies-split}, 
we have to resort to the option of the integrability breakdown provided by the mechanism of effective three-body 
elastic collisions in one dimension \cite{M089}.

This work was supported by the the FWF (Project No. P22590-N16). The authors thank J. Burgd\"orfer, 
N. J. Mauser, N. P. Proukakis, J. Schmiedmayer, and H.-P. Stimming for helpful discussions.


\begin{thebibliography}{99}
\bibitem{prim1} In our simulations we neglect the fluctuations of $\beta _k$ and always choose 
$\beta _k =\sqrt{{\cal N}_0(k)}$.
\bibitem{prim2} $T_\mathrm{eq}$ is to be determined from Eq. (\ref{eq:3}) and not from $\sum _k {\cal N}_0 = 
\sum _k {\cal N}_\mathrm{BE}$ since the total number of elementary excitations is not conserved.
\bibitem{prim3} The distribution of momenta of atoms (not to be confused with that of Bogoliubov 
quasiparticles) is given by the Fourier transform of $g_1$.
\end{thebibliography}
\end{document}